\documentclass[pra,twocolumn,amsmath,amssymb,
superscriptaddress,showpacs]{revtex4}
\usepackage{graphics}
\usepackage{epsfig}
\usepackage{bbm}

\newcommand{\ket}[1]{\ensuremath{\left|{#1}\right\rangle}}

\def\be{\begin{equation}}
\def\ee{\end{equation}}
\def\eea{\end{eqnarray}}
\def\bea{\begin{eqnarray}}

\newcommand{\exs}[1]{\ensuremath{\langle{#1}\rangle}}

\begin{document}
\title{Addendum to ``Sufficient
conditions for three-particle entanglement
 and their tests in recent experiments''}
\date{\today}

\begin{abstract}
A recent paper [M. Seevinck and J. Uffink, Phys. Rev. A {\bf
65}, 012107 (2002)] presented a bound for the three-qubit
Mermin inequality such that the violation of this bound indicates
genuine three-qubit entanglement. We show that this bound can be
improved for a specific choice of observables. In particular, if
spin observables corresponding to orthogonal directions are
measured at the qubits (e.g., X and Y spin coordinates) then the
bound is the same as the bound for states with a local hidden
variable model. As a consequence, it can straightforwardly be
shown that in the experiment described by J.-W. Pan {\it et al.} 
[Nature {\bf 403}, 515 (2000)] genuine three-qubit entanglement was
detected.
\end{abstract}

\author{G\'eza T\'oth}
\email{toth@alumni.nd.edu}
\affiliation{Max-Planck-Institut f\"ur
Quantenoptik, Hans-Kopfermann-Stra{\ss}e 1, D-85748 Garching,
Germany.}
\author{Otfried G\"uhne}
\email{otfried.guehne@uibk.ac.at}
\affiliation{Institut f\"ur
Quantenoptik und Quanteninformation, \"Osterreichische Akademie
der Wissenschaften, A-6020 Innsbruck, Austria}
\author{Michael Seevinck}
\email{m.p.seevinck@phys.uu.nl}
\affiliation{Institute of History and Foundations
of Science, Utrecht University,
P.O Box 80.000, 3508 TA Utrecht, The Netherlands}
\author{Jos Uffink}
\email{J.B.M.Uffink@phys.uu.nl}
\affiliation{Institute of History and Foundations of Science,
Utrecht University,
P.O Box 80.000, 3508 TA Utrecht, The Netherlands}

\maketitle

Before presenting our comments concerning Ref.~\cite{SU01},
let us briefly recall the difference between genuine
three-qubit entanglement, partial (biseparable) entanglement,
and full separability. A pure state is called fully
separable iff it is of the product form
$\ket{\psi}=\ket{a}\ket{b}\ket{c}.$ It is called biseparable
iff a partition of the qubits
into two groups can be found, for which the state is separable,
while for genuine three-qubit entangled states this is not
possible. For example, the state
$(\ket{000}+\ket{110})/\sqrt{2}=(\ket{00}+\ket{11})\ket{0}/\sqrt{2}$
is biseparable since the third qubit is not entangled with
the first two, while
$\ket{GHZ}=(\ket{000}+\ket{111})/\sqrt{2}$ is a state with
genuine three-qubit entanglement \cite{mixedmqubit}.

Let us now consider the three-qubit Mermin inequality. Its Bell
operator is \cite{M90}
\begin{eqnarray}
M_3&:=&X^{(1)}X^{(2)}X^{(3)}
-Y^{(1)}Y^{(2)}X^{(3)}\nonumber\\
&-&X^{(1)}Y^{(2)}Y^{(3)} -Y^{(1)}X^{(2)}Y^{(3)},\label{mermin}
\end{eqnarray}
where $X$ and $Y$ are Pauli spin matrices.
Note that these observables correspond to
orthogonal measurement directions.
For states
allowing a local hidden variable model it is known that the Mermin
inequality $|\exs{M_3}|\le 2$ has to hold \cite{M90}.
Consequently, all fully separable states have to obey the same
inequality, since they allow for a local hidden variable model.

Let us now show that for biseparable states the bound is also $2,$
thus $|\exs{M_3}| > 2$ implies that the state carries genuine tripartite
entanglement. For biseparable pure states
of the form $\Psi=\Psi_{12}\otimes\Psi_{3}$ we have
\begin{eqnarray}
\exs{M_3}&=&\exs{X^{(1)}X^{(2)}}\exs{X^{(3)}}-
\exs{Y^{(1)}Y^{(2)}}\exs{X^{(3)}}\nonumber\\
&-&\exs{X^{(1)}Y^{(2)}}\exs{Y^{(3)}}-
\exs{Y^{(1)}X^{(2)}}\exs{Y^{(3)}}.
\end{eqnarray}
Now $\exs{M_3}$ is given with six operator expectation values. We
are looking for its maximum, with the constraint that these
expectation values have physically accessible values. That is, for
example, the values for $\exs{X^{(3)}}$ and $\exs{Y^{(3)}}$ obey
$\exs{X^{(3)}}^2+\exs{Y^{(3)}}^2\le 1.$ Let us now define two
vectors with the expectation values as
\begin{eqnarray}
 \vec{v}_1&:=&(\exs{X^{(1)}X^{(2)}-Y^{(1)}Y^{(2)}},
-\exs{X^{(1)}Y^{(2)}+Y^{(1)}X^{(2)}}),\nonumber\\
 \vec{v}_2&:=&(\exs{X^{(3)}},\exs{Y^{(3)}}).
\end{eqnarray}
The relevant constraints for physical states are summarized in
$|\vec{v}_{1}|\le 2$ \cite{PROOF} and $|\vec{v}_{2}|\le 1$. One
easy way for getting an upper bound for $|\exs{M_3}|$ is using the
Cauchy-Schwarz inequality
\begin{eqnarray}
|\exs{M_3}|=|\vec{v}_{1}\cdot
\vec{v}_{2}|\le|\vec{v}_{1}||\vec{v}_{2}|\le2,
\end{eqnarray}
where $\cdot$ denotes scalar product. Since $M_3$ is invariant
under permuting qubits, this bound is clearly valid also for
biseparable pure states with partitions $(1)(23)$ and $(13)(2)$.
It is easy to see that the bound is also valid for the mixture of
pure biseparable states \cite{TG04}.

In Ref. \cite{SU01} it was shown that for biseparable quantum
states \be |\exs{M_3}|\le 2\sqrt{2} \label{uffe} \ee has to hold,
and that violation of this bound implies genuine tripartite
entanglement. This bound was found by considering the following
expression \be \Sigma:= |E(a,b,c)-E(a',b',c)-E(a,b',c')-E(a',b,c')|,
\nonumber \label{Sigma} \ee where  $a,b,c, a',b',c'$ are arbitrary
dichotomic observables on each qubit and $E(a,b,c)$ denotes the
expectation value of the corresponding product $a \cdot b\cdot c,$
if $a$ is measured at the first qubit, $b$ is at the second, etc.
The upper bound of $\Sigma$ for fully separable states is
known to be $\Sigma \leq 2.$

References
\cite{SU01,G01,CG02} show that for biseparable quantum states we have
$\Sigma \leq 2\sqrt{2}$.
Whether this bound is sharp can be checked by looking for states
for which $\Sigma = 2\sqrt{2}.$ Such a state is presented in
Ref.~\cite{SU01}: For this example $c=c'$. Thus this
bound is sharp if we allow {\it arbitrary} observables to be
measured at the qubits. When the observables are fixed to be $X$
and $Y,$ the bound can be improved, as we have seen above. The
bound in our new inequality $|\exs{M_3}| \le 2$ is sharp since for
the biseparable state $(\ket{00}+\ket{11})(\ket{0}+\ket{1})/2$
we have $|\exs{M_3}| = 2.$

In  Ref. \cite{SU01} several experiments were discussed and it was
investigated, whether three-qubit entanglement was present in
these experiments. The experiment described in Ref.~\cite{PB00}
aimed to create a GHZ state with three photons. Here the
horizontal/vertical polarization of the photons encoded the 
one-qubit information. After the state was created, the polarization
of the photons was measured. In particular, for the expectation
value of $M_3$ they obtained $2.83\pm0.09$. Based on our present
discussion, we can say, based {\it merely} on the value measured
for $\exs{M_3}$, that three-qubit entanglement has been detected
in these experiments.

In this Addendum, we have shown that the general bound obtained in
Ref. \cite{SU01} for biseparable states can be improved for the
specific choice of orthogonal observables as in the three-qubit
Mermin inequality. We determined the sharp bound for this case. We
showed that the three-qubit Mermin inequality can be used to
detect genuine three-qubit entanglement with the same bound which
is used to detect the violation of local realism. This sheds new
light on old experimental data in Ref.~\cite{PB00} and shows that
genuine three-qubit entanglement has already been realized
experimentally \cite{Nagata}. Our discussion also has a message
for using the Mermin inequality to detect genuine multiparticle
entanglement in future experiments.

We would like to thank M.~Aspelmeyer,
J.I.~Cirac, J.-W.~Pan,
H.~Weinfurter, A.~Zeilinger and M.~\.Zukowski
for useful discussions. We also
acknowledge the support of the EU project RESQ and QUPRODIS and
the Kompetenznetzwerk Quanteninformationsverarbeitung der
Bayerischen Staatsregierung. G.T. thanks the Marie Curie
Individual Fellowship (Grant No. MEIF-CT-2003-500183).



\begin{thebibliography}{99}
\bibitem{mixedmqubit} Consequently, a mixed state is called biseparable
(fully separable) if it is the mixture of pure biseparable (fully separable)
states.
\end{thebibliography}
\end{document}